\documentclass[12pt,twoside]{article}
\usepackage[mathscr]{eucal}
\usepackage{amsmath,amsfonts,amssymb,amsthm,mathabx,empheq}
\usepackage{times}
\usepackage{pdfsync}
\usepackage{cite}
\usepackage{url}
\usepackage{hyperref}
\usepackage{tensor}
\usepackage{color}
\usepackage{multicol}
\usepackage{bbold}

\advance\textheight\topskip
\allowdisplaybreaks[1]

\newcommand\td{\text{d}}
\newcommand\cO{{\cal O}}

\newcommand{\p}{\partial}
\def \sg {\sqrt{-g}}
\newcommand{\be}{\begin{equation}}
\newcommand{\ee}{\end{equation}}
\newcommand{\bea}{\begin{eqnarray}}
\newcommand{\eea}{\end{eqnarray}}
\def\nn{\nonumber}
\def\bz{\bar z}
\def\bw{\bar w}
\def\bphi{\bar \phi}
\def\ga{\gamma_{z\bz}}

\def\cH{{\cal H}}
\def\cL{{\cal L}}

\def \gai {\gamma_{z\bz}^{-1}}

\def\n{\nabla}
\def \e{ \varepsilon}

\def \ga {\gamma_{z\bz}}
\def \gai {\gamma_{z\bz}^{-1}}
\def \sg {\sqrt{-g}}

\def \e{ \varepsilon}
\def \tit {\textit}
\def \g{g_{_{YM}}}

\def\cQ{\mathcal{Q}}

\newcommand*\xbar[1]{%
  \hbox{%
    \vbox{%
      \hrule height 0.5pt % The actual bar
      \kern0.3ex%         % Distance between bar and symbol
      \hbox{%
        \kern-0.0em%      % Shortening on the left side
        \ensuremath{#1}%
        \kern-0.0em%      % Shortening on the right side
      }%
    }%
  }%
}

\DeclareFontFamily{OT1}{rsfs}{} \DeclareFontShape{OT1}{rsfs}{m}{n}{
<-7> rsfs5 <7-10> rsfs7 <10-> rsfs10}{}
\DeclareMathAlphabet{\mycal}{OT1}{rsfs}{m}{n}

\begin{document}
\title{Soft gluon theorems in curved spacetime}

\author{Peng Cheng and Pujian Mao}

\date{}

\def\mytitle{Soft gluon theorems in curved spacetime}

\addtolength{\headsep}{4pt}

\begin{centering}

  \vspace{1cm}

  \textbf{\Large{\mytitle}}

  \vspace{1.5cm}

{\large Peng Cheng$^{a,b}$ and Pujian Mao$^{a}$}%$^{\clubsuit}$$^{\diamondsuit}$

\vspace{.5cm}

\vspace{.5cm}
\begin{minipage}{.9\textwidth}\small \it  \begin{center}
    ${}^a$ Center for Joint Quantum Studies and Department of Physics,\\
     School of Science, Tianjin University, 135 Yaguan Road, Tianjin 300350, China
 \end{center}
\end{minipage}

\vspace{0.2cm}
\begin{minipage}{.9\textwidth}\small \it  \begin{center}
    ${}^b$ Lanzhou Center for Theoretical Physics,\\
    Key Laboratory of Theoretical Physics of Gansu Province,\\
    Lanzhou University, 222 South Tianshui Road, Lanzhou 730000, Gansu, China
 \end{center}
\end{minipage}

\end{centering}

%\begin{center}
%Emails:pjmao@tju.edu.cn%$^{\clubsuit}$$^{\diamondsuit}$
%\end{center}

\vspace{1cm}

\begin{center}
\begin{minipage}{.9\textwidth}
  \textsc{Abstract}. In this paper, we derive a soft gluon theorem in the near horizon region of the Schwarzschild black hole from the Ward identity of the near horizon large gauge transformation. The flat spacetime soft gluon theorem can be recovered as a limiting case of the curved spacetime.
 \end{minipage}
\end{center}
\thispagestyle{empty}

%\newpage

\section{Introduction}

Soft theorems reveal the universal properties that scattering amplitudes must obey. Such universal properties are shown in recent years to be governed by the symmetries of the S-matrix, namely the asymptotic symmetries of the theory \cite{Strominger:2013lka,Strominger:2013jfa,He:2014laa}, see also \cite{Strominger:2017zoo} for a review and references therein. The relation between soft theorem and asymptotic symmetry can be summarized as follows. Asymptotic symmetries are spontaneously broken symmetries and the soft particles are nothing but the Goldstone bosons associated to the asymptotic symmetries. Soft theorems are interpreted as the Ward identity of asymptotic symmetries. The vacuum of a theory with spontaneously broken symmetry is degenerate. Based on this point, Hawking, Perry, and Strominger (HPS) argued that black holes can carry soft hairs which are soft particles on the horizon and the soft particles are just the Goldstone bosons of asymptotic symmetries \cite{Hawking:2016msc}. In the proposal of HPS, it is directly assumed that soft theorems do exist in curved spacetime, e.g., in Schwarzschild spacetime for simplicity, and soft theorems in curved spacetime are also connected to asymptotic symmetries. Nevertheless, those assumptions are not obviously guaranteed. In particular, the previous point seems even questionable because the definition of scattering amplitudes in curved spacetime is not entirely clear until now. In a previous article \cite{Cheng:2022xyr}, we revealed a soft photon theorem in the near horizon region of the Schwarzschild black hole. Our derivation is justified by the following facts. When spacetime possesses a horizon, the horizon cuts the spacetime into two parts. Spacetime on different sides of the horizon will have different causal property and one should normally define the theory of relevance on one side of the horizon. Then the horizon should be considered as the inner boundary of the spacetime. So near horizon symmetries should have the same effect as asymptotic symmetries to the theory. Consequently, a near horizon soft theorem can be derived from the near horizon symmetry which was demonstrated precisely by Maxwell theory in \cite{Cheng:2022xyr}. Regarding to the near horizon soft theorem, two natural questions arise. The first one is that if the derivation in \cite{Cheng:2022xyr} can be extended to other theories. The other is that if the spacetime curvature characterized by the mass parameter of the Schwarzschild black hole affects the near horizon soft theorem for different theories universally. Considering the robustness of the relation between asymptotic symmetry/near horizon symmetry and soft theorem, the first question should have a definite answer. While, there is no obvious answer to the second question a priori. The aim of the present work is to touch those two points by examining the near horizon soft theorem for a non-Abelian gauge theory in the Schwarzschild spacetime.

In this paper, we study a scalar matter coupled non-Abelian gauge theory in the near horizon region of the Schwarzschild black hole. For generality, we keep the gauge group arbitrary. The scalar fields are in the adjoint representation of the gauge group. With reasonable gauge and near horizon boundary conditions, we work out the near horizon symmetries, solution space, and boundary charges on the horizon. The boundary charge on the horizon has a very similar form as the null infinity case derived in \cite{Mao:2017tey}. Meanwhile, the derivation of a soft gluon theorem from the near horizon charge needs two adaptations. The first one is to perform a mode expansion for the non-Abelian gauge fields in the Schwarzschild spacetime. The second one is to generalize the flat spacetime commutation relation between gauge fields to the near horizon region. For the first adaptation, the mode expansion for Abelian gauge fields in \cite{Cheng:2022xyr} can be extended to non-Abelian case straightforwardly. For the second adaptation, we propose a commutator between gauge fields with lower and upper spacetime index. In the new definition, the commutator of scalar fields, gauge fields, and metric fields are given in a universal way. Then, we derive a near horizon soft gluon theorem in the position space from the Ward identity of the near horizon charge. The effect from the spacetime curvature is the same as the near horizon soft photon theorem in \cite{Cheng:2022xyr}. Such result indicates that the horizon structure is universal in the context of the near horizon soft theorem. After translating to momentum space, the flat spacetime soft gluon theorem can be recovered from the large mass limit of the Schwarzschild spacetime.

The organization of this paper is very simple. In the next section, we derive the near horizon symmetry, solution space, and charge for the non-Abelian gauge theory. In section \ref{soft}, we derive a near horizon soft gluon theorem in both position space and momentum space. We close by some concluding remarks in the last section.

%%%%%%%%%%%%%%%%%%%%%%%%%%%%%%%%%%%%%%%%%%%%%%%%%%%%%%%%%%%%%%%%%%%%%%%%%%%%%%%%%%%%%%%%%%%%%%%%%%%%%%%%%%%%%%%%%%%%%%%%%%%%%%%%%%%%%%%%%%%%%%%%%%%%%%%%%%%%%%%%%%%%%%%%%%%%%%%%%%%%%%%%%%%%%%%%%%%%%%%%%%%%%%%%%%%%%%%%%%%%%%%%%%%%%%%%%%%%%%%%%%%%%%%%%%%%%%%%%%%%%%%%%%%%%%%%%%
\section{Near horizon non-Abelian gauge theory}
\label{solution}

We consider non-Abelian gauge theory in the near horizon region of the Schwarzschild black hole. The line element of the Schwarzschild black hole in the retarded coordinates $(u,r,z,\bz)$ is
\begin{equation}
\label{metric}
\td s^2=-\frac{r}{\Omega}\td u^2-2\td u\td r+2\Omega^2\gamma_{z\bz}\td z\td\bz,
\end{equation}
where
\begin{equation}
\gamma_{z\bz}=\frac{2}{(1+z\bz)^2},\quad \Omega=(r+2M).\nn
\end{equation}
Here $(z,\bz)$ are the complex stereographic coordinates. The horizon $\cH$ is just the null hypersurface $r=0$ in the retarded spherical coordinates. The non-zero components of the Levi-Civit\`{a} connection in the retarded coordinates are
\begin{align}
&\Gamma^u_{uu}=-\frac{M}{\Omega^2},\quad \Gamma^u_{z\bz}=\Omega \ga,\quad \Gamma^r_{uu}=\frac{rM}{\Omega^3},\quad \Gamma^r_{ur}=\frac{M}{\Omega^2},\quad \Gamma^r_{z\bz}=-r\ga,\nn\\ &\Gamma^z_{rz}=\frac{1}{\Omega},\quad \Gamma^z_{zz}=\p_z\ln\ga,\quad \Gamma^{\bz}_{r\bz}=\frac{1}{\Omega},\quad \Gamma^{\bz}_{\bz\bz}=\p_{\bz}\ln\ga.\nn
\end{align}

\begin{figure}[ht] % Example image
\center{\includegraphics[width=1\linewidth]{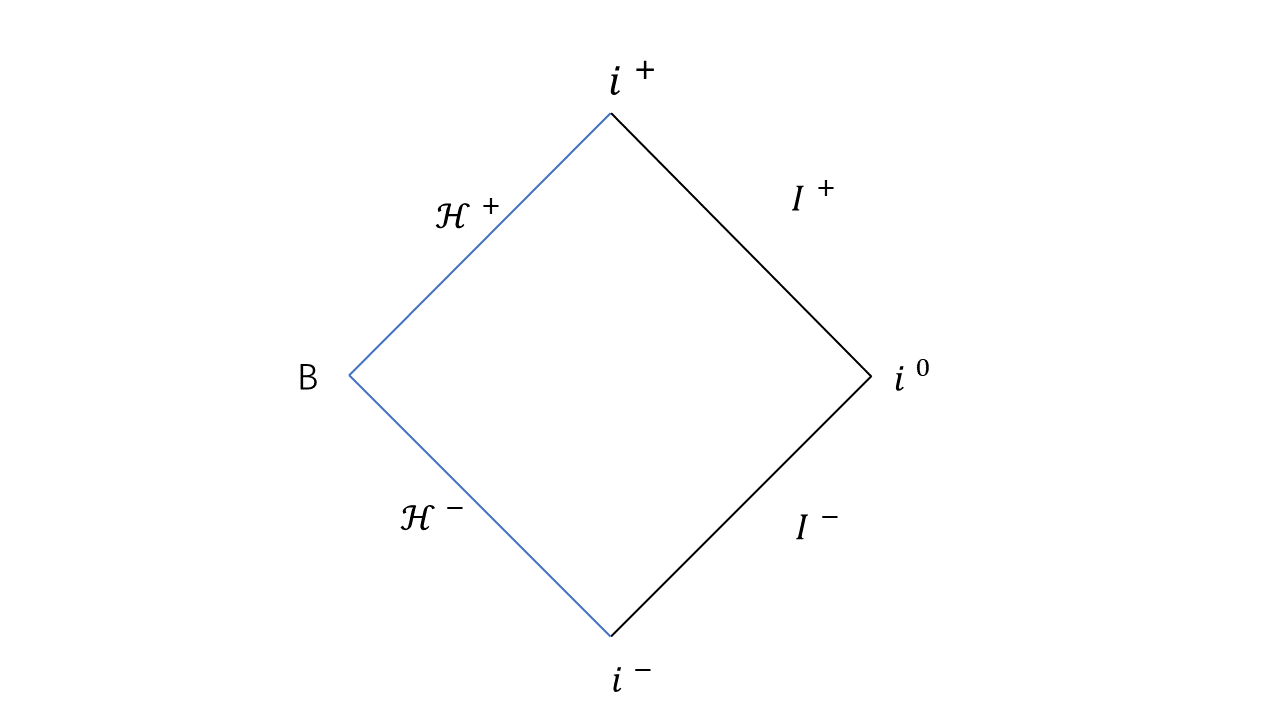}}
\caption{The Penrose diagram of the outside region of the Schwarzschild black hole.} \label{f1}
\end{figure}

In this work, we will only pay attention to half of the horizon $\cH^-$ as demonstrated in Figure \ref{f1}, because the retarded coordinates can only cover half of the horizon. The other half $\cH^+$ can be studied in a similar way in the advanced coordinates $(v,r,z,\bz)$. The near horizon symmetries on $\cH^-$ and $\cH^+$ can be connected by the symmetries near the bifurcation point $B$ \cite{Adami:2020amw}.

For simplicity, we consider a non-Abelian gauge theory with only scalar matter in the adjoint representation of the gauge group $G$. The extension to matter fields in the fundamental representation or fermionic matter should be straightforward. The Lagrangian of the non-Abelian gauge theory is \cite{Mao:2017tey}
\be\label{lagrangian}
\cL=\frac14F_{\mu\nu}^aF^{a\mu\nu} + D_\mu\phi^a (D^\mu\phi^a)^\dag,
\ee
where the field-strength is given by
\be
F_{\mu\nu}^a=\p_\mu A_\nu^a - \p_\nu A_\mu^a +i\g C^{abc}A^b_\mu A^c_\nu,
\ee
and the covariant derivative associated to the gauge group $D_\mu$ is defined as
\be
D_\mu X^a=\n_\mu X^a + i\g C^{abc} A^b_\mu X^c.
\ee
The group structure constant $C_{abc}$ is totally antisymmetric. The structure constant satisfies the Jacobi identity,
\be
C^{abc}C^{bde} + C^{abd}C^{bec} + C^{abe}C^{bcd}=0.
\ee
One can derive the equations of motion from the Lagrangian \eqref{lagrangian} as
\be\label{eom}
D_\mu F^{c\mu\nu}=\g J^{c\nu},\quad D_\mu D^\mu\phi^a=0,\quad D_\mu D^\mu\bphi^a=0,
\ee
where
\be\label{current}
J^{c\nu}=i C^{abc}(\phi^a D^\nu\bphi^b - \bphi^b D^\nu \phi^a).
\ee
The Lagrangian \eqref{lagrangian} is invariant up to the total derivative under the gauge transformation
\be\label{gaugetransf}
\delta_\e A_\mu^a=D_\mu \e^a,\quad \delta_\e \phi^a=i\g C^{abc}\phi^b\e^c.
\ee
We will work with the radial gauge,
\be\label{condition}
A_r^a=0.
\ee
The near horizon conditions of the rest components of the gauge fields and the scalar fields are chosen as
\be\label{boundary}
A_u^a=\cO(r),\quad A_{z}^a=\cO(1),\quad A_{\bz}^a=\cO(1),\quad \phi^a=\cO(1),\quad \bphi^a=\cO(1).
\ee
The gauge and near horizon conditions yield the residual gauge transformation parameter $\e^a(z,\bz)$. One can further check that the residual gauge transformation satisfies the symmetry algebra
\be
[\delta_{\e_{1}},\delta_{\e_{2}}]A_\mu^a=\delta_{[\e_{1},\e_2]}A_\mu^a,\quad [\delta_{\e_{1}},\delta_{\e_{2}}]\phi^a=\delta_{[\e_{1},\e_2]}\phi^a.
\ee
For a gauge theory, the equations of motion in \eqref{eom} are connected off-shell by the Noether identity\footnote{The Noether identity of this theory can be written in a more elegant way, $D_\nu D_\mu F^{c\mu\nu} =0$. In practice, we find that \eqref{Bianchi} is more convenient.}
\be\begin{split}\label{Bianchi}
D_\nu\left[D_\mu F^{c\mu\nu}-\g J^{c\nu}\right]=i\g C^{abc}\left[\bphi^b D_\mu D^\mu\phi^a - \phi^a D_\mu D^\mu\bphi^b \right].
\end{split}\ee
The equations of motions will be arranged into four types \cite{Mao:2017tey}, namely
\begin{itemize}
\item Hypersurface equations:
\be
D_\mu F^{a\mu u}=\g J^{au}.
\ee
\item Standard equations:
\be
D_\mu F^{a\mu z}=\g J^{az},\quad D_\mu F^{a\mu \bz}=\g J^{a\bz}, \quad D_\mu D^\mu\phi^a=0,\quad D_\mu D^\mu\bphi^a=0.
\ee
\item Supplementary equations:
\be
D_\mu F^{a\mu r}=\g J^{ar}.
\ee
\end{itemize}
Once the hypersurface equations and the standard equations are satisfied, the Noether identity \eqref{Bianchi} will be reduced to
\be
\p_r\left[\sg \big( D_\mu F^{a\mu r} - \g J^{ar}\big)\right]=0,
\ee
Thus, one just needs to solve the supplementary equations
\be
\sg (D_\mu F^{a\mu r}- \g J^{ar})=0,
\ee
at order $\cO(1)$ and all the other orders (in $r$) are zero because of the Noether identity.

We start to solve the equations of motion from the hypersurface equations $D_\mu F^{a\mu u}=\g J^{au}$, which are reduced to
\begin{multline}
\p_r\left(\Omega^2\ga \p_rA_u^a\right)=\p_z\p_r A_{\bz}^a+\p_{\bz}\p_r A_z^a\\
+ i\g C^{abc}\left[A^b_z \p_r A_{\bz}^c + A_{\bz}^a \p_r A^c_z + \Omega^2\ga (\phi^b\p_r \bphi^c - \bphi^c\p_r \phi^b)\right].
\end{multline}
The solutions of the hypersurface equations with respect to the boundary conditions are
\begin{multline}
A_u^a=\frac{A_u^{a(0)}(u,z,\bz)}{2M}-\frac{A_u^{a(0)}(u,z,\bz)}{\Omega}+\gai\int^\infty_r \frac{\td r'}{\Omega^2}\int^\infty_{r'} \td r'' \big[\p_z \p_{r''} A_{\bz}^a + \p_{\bz} \p_{r''} A_{z}^a \\
+i\g C^{abc}\big(A^b_z \p_{r''} A^c_{\bz} + A^b_{\bz} \p_{r''} A^c_{z} + {\Omega}^2\ga \phi^b \p_{r''} \bphi^c - {\Omega}^2\ga\bphi^c \p_{r''} \phi^b \big) \big],
\end{multline}
where $A_u^{a(0)}(u,z,\bz)$ are integration constants. Suppose that $A_{z}^a$, $A_{\bz}^a$, $\phi^a$, and $\bphi^a$ are given in series expansion as
\be\begin{split}
&A_{z}^a=\sum\limits_{m=0}^\infty A_{z}^{a(m)}(u_0,z,\bz) r^{m} ,\quad A_{\bz}^a=\sum\limits_{m=0}^\infty A_{\bz}^{a(m)}(u_0,z,\bz) r^{m} ,\\
&\phi^a=\sum\limits_{m=0}^\infty \phi^{a(m)}(u_0,z,\bz) r^{m} ,\quad \bphi^a=\sum\limits_{m=0}^\infty \bphi^{a(m)}(u_0,z,\bz) r^{m},
\end{split}\ee
as initial data. Then, $A_u^a$ will be fixed up to the integration constants,
\be\label{Au}
A_u^a=\frac{A_u^{a(0)}r}{4M^2}+\cO({r^2}),
\ee
where we omit all the fixed orders which do not contain new information about the solution space.

The four groups of standard equations can be written as
\begin{multline}\label{puAz}
2\p_u\p_r A_{z}^a - \p_r\p_{z} A_u^a + i\g C^{abc} \p_r A_u^a A_{z}^c + 2 i \g C^{abc} A_u^b \p_r A_{z}^c - \p_r \left(\frac{r}{\Omega} \p_r A_{z}^a\right)\\
-\p_{z} {F^{az}}_{z} - i \g C^{abc} A_{z}^b {F^{cz}}_{z} + \g J^a_{z}=0,
\end{multline}
\begin{multline}\label{puAzb}
2\p_u\p_r A_{\bz}^a - \p_r\p_{\bz} A_u^a + i\g C^{abc} \p_r A_u^a A_{\bz}^c + 2 i \g C^{abc} A_u^b \p_r A_{\bz}^c - \p_r \left(\frac{r}{\Omega} \p_r A_{\bz}^a\right)\\
-\p_{\bz} {F^{a\bz}}_{\bz} - i \g C^{abc} A_{\bz}^b {F^{c\bz}}_{\bz} + \g J^a_{\bz}=0,
\end{multline}
\begin{multline}\label{puphi}
2\p_u \p_r \left(\Omega \phi^a\right) -\frac{1}{\Omega}\p_r \left(r\Omega \p_r \phi^a\right) - \Omega \n_z \n^z \phi^a - \Omega \n_{\bz}\n^{\bz} \phi^a \\
+i \g C^{abc} \bigg[2 A_u^b \phi^c
+ \Omega \p_r \left(A_u^b \phi^c \right) + \Omega A_u^b D_r \phi^c\\
- \Omega \n_z \left(A^{bz} \phi^c\right) - \Omega \n_{\bz} \left(A^{b\bz} \phi^c\right) - \Omega A_z^b D^z \phi^c - \Omega A_{\bz}^b D^{\bz} \phi^c\bigg]=0,
\end{multline}
\begin{multline}\label{pubphi}
2\p_u \p_r \left(\Omega \bphi^a\right) -\frac{1}{\Omega}\p_r \left(r\Omega \p_r \bphi^a\right) - \Omega \n_z \n^z \bphi^a - \Omega \n_{\bz}\n^{\bz} \bphi^a\\
+i \g C^{abc} \bigg[2 A_u^b \bphi^c
+ \Omega \p_r \left(A_u^b \bphi^c \right) + \Omega A_u^b D_r \bphi^c\\
- \Omega \n_z \left(A^{bz} \bphi^c\right) - \Omega \n_{\bz} \left(A^{b\bz} \bphi^c\right) - \Omega A_z^b D^z \bphi^c - \Omega A_{\bz}^b D^{\bz} \bphi^c\bigg]=0,
\end{multline}
which will determine the time dependence of $A_{z}^{a(m)}$, $A_{\bz}^{a(m)}$, $\phi^{a(m)}$, and $\bphi^{a(m)}$ recursively. However, the leading terms $A_z^{a(0)}$, $A_{\bz}^{a(0)}$, $\phi^{a(0)}$, and $\bphi^{a(0)}$ are not constrained. Thus, we would refer to the time derivative of them as \tit{news} functions following the terminology of \cite{Bondi:1962px}, which represent the local propagating degree of freedom.

In the end, the supplementary equations
$D_\mu F^{a\mu r}=\g J^{ar}$ control the time evolution of the integration constants as
\begin{multline}\label{puAu0}
\p_u A_u^{a(0)}=-\gai \p_u\big(\p_z A_{\bz}^{a(0)} + \p_{\bz} A_z^{a(0)}\big)\\
- i \gai \g C^{abc}\big(A_z^{b(0)}\p_u A_{\bz}^{c(0)} -  A_{\bz}^{c(0)}\p_u A_z^{b(0)}\big)\\
- i 4 M^2 \g C^{abc}\big(\phi^{b(0)}\p_u \bphi^{c(0)} -  \bphi^{c(0)}\p_u \phi^{b(0)}\big).
\end{multline}

The boundary charge for scalar matter coupled theory \eqref{lagrangian} is given by \cite{Mao:2017tey}
\be
\cQ_{\e}=\int\td z\td\bz\,\Omega^2\,\ga\,\e^a \,F^{aur}.
\ee
We define the charge at the bifurcation point on the horizon as
\be
\cQ_{\e}=\int_B \td z\td\bz\,\ga\,\e^a \, A_u^{a(0)}=\int_{\cH^-} \td u \td z\td\bz\,\ga\,\e^a \,\p_u A_u^{a(0)},
\ee
where we assumed the conditions $A_u^{a(0)}=0$ at $i^-$ in the second equality. Inserting the relation \eqref{puAu0}, the charge becomes
\be\begin{split}\label{charge}
\cQ_{\e}=-&\int_{\cH^-} \td u \td z\td\bz\,\e^a\bigg[ \p_u\big(\p_z A_{\bz}^{a(0)} + \p_{\bz} A_z^{a(0)}\big)\\
&+ i  \g C^{abc}\big(A_z^{b(0)}\p_u A_{\bz}^{c(0)} -  A_{\bz}^{c(0)}\p_u A_z^{b(0)}\big)\\
&+ i 4  \ga M^2 \g C^{abc}\big(\phi^{b(0)}\p_u \bphi^{c(0)} -  \bphi^{c(0)}\p_u \phi^{b(0)}\big)\bigg].
\end{split}\ee

%%%%%%%%%%%%%%%%%%%%%%%%%%%%%%%%%%%%%%%%%%%%%%%%%%%%%%%%%%%%%%%%%%%%%%%%%%%%%%%%%%%%%%%%%%%%%%%%%%%%%%%%%%%%%%%%%%%%%%%%%%%%%%%%%%%%%%%%%%%%%%%%%%%%%%%%%%%%%%%%%%%%%%%%%%%%%%%%%%%%%%%%%%%%%%%%%%%%%%%%%%%%%%%%%%%%%%%%%%%%%%%%%%%%%%%%%%%%%%%%%%%%%%%%%%%%%%%%%%%%%%%%%%%%%%%%%%%
\section{Near horizon soft gluon theorem}
\label{soft}

The crucial point to recover a soft theorem from asymptotic symmetry is that the charge associated to a spontaneously broken symmetry acts non-linearly on the quantum states. Then, one can, in principle, split the charge into linear and non-linear pieces $Q=Q_{\rm{L}}+Q_{\rm{NL}}$. The corresponding Ward identity can be written as
\begin{equation}
\label{Wards}
	\langle\rm{out}|Q_{\rm{NL}}^{\textrm{out}}-Q^{\textrm{in}}_{\rm{NL}}|\rm{in}\rangle=
	-\langle\rm{out}|Q^{\textrm{out}}_{\rm{L}}-Q^{\textrm{in}}_{\rm{L}}|\rm{in}\rangle.
\end{equation}
The near horizon symmetry does not preserve the vacuum of the theory, it must be spontaneously broken. Let us now decompose the charge~\eqref{charge} as follows:\footnote{We drop the total minus sign in the charge which will not affect the Ward identity.}
\begin{align}
\label{Q0NLm}
	Q_{\textrm{NL}}=&\int_{\cH^-}\td z\td\bz \td u\,\varepsilon^a\,\partial_u\partial_{\bz}A^{a(0)}_z,\\
\label{Q0Lm}
Q_{\textrm{L}}=&\frac12\int_{\cH^-}\td z\td\bz \td u\,i  \g C^{abc}\bigg[\big(A_z^{b(0)}\p_u A_{\bz}^{c(0)} -  A_{\bz}^{c(0)}\p_u A_z^{b(0)}\big)\nn\\
&+  4  \ga M^2 \big(\phi^{b(0)}\p_u \bphi^{c(0)} -  \bphi^{c(0)}\p_u \phi^{b(0)}\big)\bigg].
\end{align}
These are, respectively, the non-linear and linear pieces of the charge for near horizon symmetry. We keep only the anti-holomorphic terms for the non-linear charge and split the linear charge into two parts. That is the reason for the factor $\frac12$ in \eqref{Q0Lm}. Such treatment is consistent with the Abelian case in~\cite{He:2014cra} where the authors consider only the sector of the phase space with no long-range magnetic fields, in which case, the helicity of soft particle is decoupled and one can only deal with either the holomorphic part or the anti-holomorphic part.

In the language of S-matrix, a symmetry acts on both the \tit{in} and \tit{out} states which defined on $\cH^-$ and $\cH^+$ respectively. The previous analysis for the near horizon charge is carried out only near $\cH^-$. But it can be performed similarly on $\cH^+$ and an identification at the bifurcation point can be set to define a full charge for S-matrix. So we will only deal with the \tit{in} part in the present work and, for notational brevity, we suppress the \tit{in} label on the states.

The anti-holomorphic soft gluon theorem is obtained from the Ward identity by a concrete choice of $\varepsilon^a(z,\bz)$ at null infinity~\cite{Mao:2017tey}. The same choice will be adopted for the near horizon case,
\begin{equation}
\label{eps}
	\varepsilon^a(z,\bz)=\frac{1}{w-z},
\end{equation}
from which, one has $\partial_{\bz}\varepsilon^a=-2\pi\delta^2(z-w)$.

\subsection{Mode expansion in Schwarzschild spacetime}

The mode expansion of an Abelian gauge field in the Schwarzschild background was derived in \cite{Cheng:2022xyr}. The extension to non-Abelian case is straightforward. Correspondingly, the action of the non-linear charge on the states is very similar to the Abelian case. For completeness, we briefly present the derivation. The main difference for the mode expansion in curved spacetime is the dispersion relation of the momentum of a moving particle in curved spacetime which can be adapted in the isotropic coordinates system $(t,x_i)$, where the line element of the Schwarzschild black hole is
\be\label{isoflat}
\td s^2=-\left(\frac{2\rho-M}{2\rho+M}\right)^2\td t^2 +\frac{(2\rho+M)^4}{16\rho^4}d\vec x^2\,.
\ee
Now the horizon is located at $\rho=\frac{M}{2}$. The dispersion relation for the massless particle in the isotropic coordinates is
\be\label{dispersion}
-\left(\frac{2\rho+M}{2\rho-M}\right)^2 \omega^2+\frac{16\rho^4}{(2\rho+M)^4}{\vec p~}^2=0,
\ee
where $\vec p$ is the three momenta of the conformally flat part. For a free massless scalar field $\Phi(x)$, we can write the field operator as \cite{Mukhanov:2007zz}
\begin{equation}
	\Phi(x^\mu)=-\frac{1}{(2\pi)^4}\int
	\td\omega \td^3\vec p~\tilde{\phi}(p_\mu)~e^{ip\cdot x}.\label{expansion}
\end{equation}
Inserting the dispersion relation as a delta function in the integral yields
\begin{eqnarray*}
\Phi(x^\mu) &=& \frac{1}{(2\pi)^3}\left(\frac{2\rho-M}{2\rho+M}\right)^2\int \frac{\td^3\vec p}{2\omega}~\tilde{\phi}(p_\mu)e^{ip\cdot x}\Big{|}_{-p_0=\omega}\\
&=&\frac{1}{(2\pi)^3}\left(\frac{2\rho-M}{2\rho+M}\right)^2\int ~\frac{\td^3\vec p}{2\omega}~\left[\mathfrak{a}(p_\mu)e^{ip\cdot x}+\mathfrak{a}^{\dagger}(p_\mu)e^{-ip\cdot x}\right]\Big{|}_{\omega>0}.
\end{eqnarray*}
We have an extra factor $\frac{({2\rho-M})^2}{({2\rho+M})^2}$ for the mode expansion because of the dispersion relation \eqref{dispersion}.

For a non-Abelian gauge field $A_{\mu}^a$, the mode expansion includes two polarization vectors orthogonal to the propagating direction,
\begin{equation}
A_{\mu}^a(x)=\sum_{\alpha=\pm}\frac{1}{(2\pi)^3}\left(\frac{2\rho-M}{2\rho+M}\right)^2 \int ~\frac{\td^3\vec p}{2\omega}~\left[\epsilon_\mu^{*\alpha} \mathfrak{a}^a_{\alpha}(p)e^{ip\cdot x} + \epsilon_\mu^{\alpha}\mathfrak{a}_{\alpha}^{a\dagger}(p)e^{-ip\cdot x}\right],
\end{equation}
where the polarization vectors satisfy $\epsilon_{\alpha}^{\mu}\epsilon^*_{\beta\mu}=\delta_{\alpha\beta}$. It is convenient to
parametrize the gluon four-momenta as
\be\label{softp}
p_{\mu}= \frac{\omega}{1+z\bar z}\frac{(2\rho+M)^3}{4 \rho^2 (2\rho-M)}\left(\frac{4 \rho^2 (M-2\rho)}{(2\rho+M)^3}(1+z\bar z), (z+\bar z),-i(z-\bar z),(1-z\bar z)\right),
\ee
and to parametrize the polarization vectors as
\be\label{polarization}
\begin{split}
\epsilon^{+\mu}=& \frac{1}{\sqrt{2}}\frac{4\rho^2}{(2\rho+M)^2}\left(\frac{(2\rho+M)^3}{4\rho^2(2\rho-M)}\bar{z},1,-i,-\bar z\right),\\
\epsilon^{-\mu}=& \frac{1}{\sqrt{2}}\frac{4\rho^2}{(2\rho+M)^2}\left(\frac{(2\rho+M)^3}{4\rho^2(2\rho-M)}z,1,i,-z\right).
\end{split}\ee
Translating to $(z,\bar z)$ coordinates, we have
\be
\epsilon^{+}_{z}=\frac{(M+2\rho)^2}{2\sqrt{2}\rho(1+z\bar z)},\quad \epsilon^{-}_{\bar z}=\frac{(M+2\rho)^2}{2\sqrt{2}\rho(1+z\bar z)}.
\ee
Inserting those expressions, the mode expansion is reduced to
\be
\begin{split}
A_{\mu}^a =& \sum_{\alpha=\pm}
\frac{(2\rho+M)^7}{(2\rho)^6(2\rho-M)}\int \frac{\omega~\td \omega}{8\pi^2}\int_0^{\pi}\td \Theta \\
&\times \left(\sin \Theta  \epsilon^{\alpha}_{\mu}\mathfrak{a}^a_\alpha e^{-i\omega t+i\omega \frac{(2\rho+M)^3}{4\rho(2\rho-M)}\cos \Theta}+c.c.\right),
\end{split}
\ee
where $\Theta$ is the angle between $\vec p$ and $\vec x$.
Let $R=\rho - \frac{M}{2}$, the $\Theta$ integration can be worked out directly using the relation
\be
\int_0^{\pi}\td \Theta \sin \Theta ~e^{i\omega \frac{2 (M+R)^3}{R (M+2 R)}\cos \Theta}=
 \frac{R (M+2 R) }{\omega  (M+R)^3}\sin \left[\frac{2 \omega  (M+R)^3}{R (M+2 R)}\right] .
\ee
In terms of $R$, the mode expansion can be written as
\begin{multline}
A_{\mu}^a
= \sum_{\alpha=\pm}
\frac{8(M+R)^4}{\pi^2(M+2R)^5}\\
 \times \int \td \omega
~ \sin \left[\frac{2 \omega  (M+R)^3}{R (M+2 R)}\right]\left(\frac{R^2}{M^2+2 M R}\right)^{-2 i M \omega }\\
\times\left( \epsilon^{\alpha}_{\mu}\mathfrak{a}^a_\alpha e^{-i\omega u-i\omega \frac{2 (M+R)^2}{M+2 R}}+c.c.\right) ,
\end{multline}
where we have adopted the retarded time $u$. Finally,
the near horizon field $A_z^{a(0)}$ is related to the plane wave modes by
\begin{multline}
A_z^{a(0)}=\frac{16\sqrt{2}}{\pi^2(1+ z \bar z)}\frac{(M+R)^6}{(M+2R)^6} \\
\times  \int \td \omega~ \sin \left[\frac{2 \omega  (M+R)^3}{R (M+2 R)}\right]\left(\frac{R^2}{M^2+2 M R}\right)^{-2 i M \omega }\\
\times\left( \mathfrak{a}^a_{+} e^{-i\omega u-i\omega \frac{2 (M+R)^2}{M+2 R}}
+\mathfrak{a}^{a\dagger}_{-} e^{i\omega u+i\omega \frac{2 (M+R)^2}{M+2 R}} \right).\label{modefinal}
\end{multline}

\subsection{Soft gluon theorem in position space}

The non-linear part of the charge \eqref{Q0NLm} can be expressed as
\be
Q_{\textrm{NL}}=\lim_{R\to 0} \int_{\cH^-}\td z\td\bz \td u\,\varepsilon^a(z,\bz)\,\partial_u \partial_{\bz}A_{z}^{a(0)} \,.
\ee
We will use the Fourier relation
\begin{equation}
	\int_{-\infty}^{\infty}\td u\,\partial_uF(u)=2\pi i\lim_{\omega\to0}\left[\omega\tilde{F}(\omega)\right],
\end{equation}
where $F(u)=\int_{-\infty}^{\infty}\td\omega\,e^{i\omega u}\tilde{F}(\omega)$.
Inserting \eqref{eps}, the non-linear charge with the mode expansion \eqref{modefinal} is reduced to
\bea
Q_{\textrm{NL}}
&=& \frac{-128\sqrt{2}i}{(1+ w \bw)} \frac{M^2}{R} \lim_{\omega\to 0}
~ \left[ \omega^2\mathfrak{a}^a_{+}
+\omega^2\mathfrak{a}^{a\dagger}_{-} \right].
\eea
Thus, the insertion of this piece of the charge in the S-matrix yields
\bea\label{QNL}
\langle\text{out}|Q_{\rm{NL}}|\text{in}\rangle
= \frac{-128\sqrt{2}i}{(1+ w \bw)} \frac{M^2}{R} \lim_{\omega\to 0}
~ \langle\text{out}| \omega^2\mathfrak{a}^{a\dagger}_{-} |\text{in}\rangle.
\eea

Regarding the action of the linear charge \eqref{Q0Lm}, one has to define canonical commutation relations for the fields at the horizon. The non-zero ones are
\be\begin{split}\label{commutator}
  [A^{a(0)}_{\bz}(u,z,\bz),A^{b\bz(0)}(u',w_k,\bw_k)]&=\frac{i}{4} \,\delta^{ab}\Omega^{-2}\,\gai\, \Theta(u'-u)\delta^2(z-w_k), \\
  [\xbar\phi^{a(0)}(u,z,\bz),\phi^{b(0)}(u',w_k,\bw_k)]&=\frac{i}{4} \,\delta^{ab}\Omega^{-2}\,\gai\, \Theta(u'-u)\delta^2(z-w_k).
\end{split}\ee
Here we define the commutator for the gauge field between a lower index one and an upper index one. The reason is the following. Essentially, the commutation relations of the fields are from the commutation relation between the creation and annihilation operators. The mode expansion of the gauge fields involves polarization vectors. The commutator of gauge fields should be with respect to the normalization relation of the polarization vectors which, on the horizon or at the null infinity, is reduced to
\be
\epsilon^{*z}_+ \epsilon_{+z}= \epsilon^{*\bz}_- \epsilon_{-\bz}=1.
\ee
Hence, we propose to define the commutator for gauge fields as in \eqref{commutator}. In such a way, one should see no difference for the commutator of scalar fields and the commutator of gauge fields. Then, one can lower the index by the sphere metric to recover the commonly used one where the commutator for gauge fields does not have a factor $\gai$, see, for instance in \cite{He:2014cra,Mao:2017tey}. Correspondingly, the commutator for metric fields should have a factor $\ga$ \cite{He:2014laa}, because one needs to lower the index twice. On the horizon, the sphere metric has a conformal factor $4M^2$ which yields
\begin{align}
  [A^{a(0)}_{\bz}(u,z,\bz),A_z^{b(0)}(u',w_k,\bw_k)]&=\frac{i}{4}  \,\delta^{ab}\, \Theta(u'-u)\delta^2(z-w_k),\label{Acommutator} \\
  [\xbar\phi^{a(0)}(u,z,\bz),\phi^{b(0)}(u',w_k,\bw_k)]&=\frac{i}{4} \,\delta^{ab}\Omega^{-2}\,\gai\, \Theta(u'-u)\delta^2(z-w_k).\label{phicommutator}
\end{align}
Defining the Fourier modes as
\be\begin{split}
&A_{E_k}^d(w_k,\bw_k)=\int \td u' e^{iE_k u'} A^{d(0)}_z(u',w_k,\bw_k),\\
&\phi_{E_k}^d(w_k,\bw_k)=\int \td u' e^{iE_k u'} \phi^{d(0)}(u',w_k,\bw_k),
\end{split}\ee
the actions of the linear charge on the Fourier modes are
\begin{align}
&[Q_{\textrm{L}},A_{E_k}^d(w_k,\bw_k)]=\frac{\g C_k^{abd}}{4(w-w_k)} A^b_{E_k}(w_k,\bw_k),\\
&[Q_{\textrm{L}},\phi_{E_k}^d(w_k,\bw_k)]=\frac{\g C_k^{abd}}{4(w-w_k)} \phi^b_{E_k}(w_k,\bw_k),
\end{align}
where the special choice of $\e^a$ in \eqref{eps} has been used. Hence, the linear charge yields
\be
\label{QL}
  \langle\textrm{out}|Q_{\textrm{L}}|\textrm{in}\rangle=\sum_{k=1}^n
  \frac{\g (T_k^a)_{i_k j_k}}{4(w-w_k)}\langle\textrm{out}|\textrm{in}\rangle_{i_1\cdots j_k \cdots i_n},
\ee
where $(T_k^a)_{i_kj_k}= C_k^{ai_kj_k}$ in adjoint representation and the index $k$ denotes the $k$th particle. Inserting \eqref{QL} and \eqref{QNL} into the Ward identity, we obtain the near horizon soft gluon theorem in asymptotic position space,
\be\label{presoftT}
\lim_{\omega\to 0}
~ \langle\text{out}| \mathfrak{a}^{\dagger}_{-}|\text{in}\rangle =
\frac{1}{512}\frac{R}{i\omega M^2}\left(\frac{1+ w \bw}{\sqrt{2}\omega}
\sum_{k=1}^n\frac{\g (T_k^a)_{i_kj_k}}{w-w_k}\right)
\langle\textrm{out}|\textrm{in}\rangle.
\ee
It is amusing to see that the terms in the parenthesis is simply the soft gluon fact in flat spacetime in the position space and the prefactor is the same as the soft photon theorem in the near horizon region of the Schwarzschild black hole \cite{Cheng:2022xyr}. Note that one should think of the Feynman prescription to regularize the radial coordinate $R$ on the horizon because the soft energy in the denominator also takes zero limit. More intuitively, one does not need any regularization for the case of null infinity because the null infinity is a conformal object. Any divergence coming from the limit $r\to\infty$ can be absorbed into the conformal factor. While, the horizon is a physical entity that particles can reach and cross it. So one should regularize the divergence from the horizon, i.e., set $R=i\tau$ and $\tau$ being infinitesimal. %From the soft factor, one can not see any effect of the black hole that is characterized by its mass parameter $M$. This may not be a surprising point considering that we derive the soft theorem in position space. The soft factor is a function of the local data of hard particles who can not see the spacetime curvature locally due to the equivalent principle. Nevertheless, the dispersion relation of the null momenta involve $M$ and the soft factor in momentum space should have the mass parameter $M$ dependence.

\subsection{Soft gluon theorem in momentum space}

Similar to the soft momentum \eqref{softp}, the hard momenta in the near horizon region can be parametrized as
\be
q_{k\mu} =\frac{E_k}{1+z_k\bar z_k}\frac{4M}{R}
\left(-\frac{R}{4M}(1+z_k\bar z_k), (z_k+\bar z_k),-i(z_k-\bar z_k),(1-z_k\bar z_k)\right).
\ee
Then, using the relations in \eqref{softp} and \eqref{polarization}, one can verify that
\be\label{useful}
\sum_{k=1}^{n} \frac{\g (T_k^a)_{i_kj_k} q_k\cdot \epsilon}{q_k\cdot p}=\frac{R}{\omega M}
\left(\frac{1+z\bz}{\sqrt{2}}\sum_{k=1}^{n} \frac{\g (T_k^a)_{i_kj_k} }{z-z_k}\right).
\ee
Inserting this relation to the soft factor in the position space, we obtain the soft gluon theorem in momentum space as
\be\label{softmomentum}
\lim_{\omega\to 0}
~ \langle\text{out}| \mathfrak{a}^{\dagger}_{-}|\text{in}\rangle=
\frac{1}{512}\frac{1}{i\omega M} \sum_{k=1}^{n} \frac{\g (T_k^a)_{i_kj_k} q_k\cdot \epsilon}{q_k\cdot p}\langle\textrm{out}|\textrm{in}\rangle .
\ee
For the soft factor in the momentum space, we should take another regularization when taking the zero mass limit ($M\to0$) of the black hole. The reason is the following. When $M=0$, we are back to the flat spacetime where $R=0$ is just a point rather than a three dimensional null hypersurface. In other words, the horizon will be shrunk to a point where one cannot even define the S-matrix or the scattering amplitude. By introducing the regularization $M\to M+i m$ and $m$ being infinitesimal, one can take the zero mass limit. On the other hand, the flat spacetime soft gluon factor can be recovered by taking the large mass limit and setting $512\omega M=1$. Technically, a large mass limit requires a large radius to compensate which makes the near horizon analysis close to the null infinity analysis. Hence, it is reasonable to see the flat space result is from the large mass limit, because the flat space soft theorem is derived from asymptotic symmetry at null infinity.

%%%%%%%%%%%%%%%%%%%%%%%%%%%%%%%%%%%%%%%%%%%%%%%%%%%%%%%%%%%%%%%%%%%%%%%%%%%%%%%%%%%%%%%%%%%%%%%%%%%%%%%%%%%%%%%%%%%%%%%%%%%%%%%%%%%%%%%%%%%%%%%%%%%%%%%%%%%%%%%%%%%%%%%%%%%%%%%%%%%%%%%%%%%%%%%%%%%%%%%%%%%%%%%%%%%%%%%%%%%%%%%%%%%%%%%%%%%%%%%%%%%%%%%%%%%%%%%%%%%%%%%%%%%%%%%%%%%
\section{Conclusion and discussion}

A soft gluon theorem in the near horizon region of the Schwarzschild black hole is derived in this paper. The soft factor consists of the flat spacetime soft factor and a prefactor $\frac{1}{\omega M}$. The inverse of the soft energy comes from the near horizon condition. It is definitely reasonable to have the black hole mass term in the soft factor since it is the only parameter that characterizes the spacetime curvature. The inverse power can be understood from the limit of recovering the flat space result.

For future directions, one would naturally consider a near horizon soft graviton theorem given the fact that the gravitational dynamics including the boundary charges on the horizon has been shown a very similar structure as the null infinity \cite{Adami:2021nnf}. But the main obstacle is again from the mode expansion. As shown in \cite{Cheng:2022xyr} and briefed in the previous section, the mode expansion has strong dependency on an isotropic coordinates which does not exist for a generic near horizon solution of full Einstein theory. Of course one way to go is to linearize the theory around a curved background, e.g., the Schwarzschild spacetime. Then, the main task is to first formulate the classical relations in analog with \cite{Adami:2021nnf} for the linearized gravity theory.

%%%%%%%%%%%%%%%%%%%%%%%%%%%%%%%%%%%%%%%%%%%%%%%%%%%%%%%%%%%%%%%%%%%%%%%%%%%%%%%%%%%%%%%%%%%%%%%%%%%%%%%%%%%%%%%%%%%%%%%%%%%%%%%%%%%%%%%%%%%%%%%%%%%%%%%%%%%%%%%%%%%%%%%%%%%%%%%%%%%%%%%%%%%%%%%%%%%%%%%%%%%%%%%%%%%%%%%%%%%

\section*{Acknowledgments}

The authors thank Laura Donnay, Zhengwen Liu, Shahin
Sheikh-Jabbari, and Jun-Bao Wu for useful discussions. The authors thank Kai-Yu Zhang for pointing out the mistake in the commutation relation. This work is supported in part by the National Natural Science Foundation of China (NSFC) under Grants No. 11905156 and No. 11935009. P.C. is also supported in part by NSFC Grant No. 12047501.

\bibliography{ref}

\end{document}